# Flowshop Machine Scheduling: Markov Modeling, Optimal Schedules and Heuristics

Samah A. M. Ghanem

*Abstract*—Flowshop machine scheduling has been of main interest in several applications where the timing of its processes plays a fundamental role in the utilization of system resources. Addressing the optimal sequencing of the jobs when equivalent failures across machines exist is a decision of particular relevance to the general scheduling problem. Such failures allow for unpredictable time consumption and improper utilization of the machines. Therefore, it is of particular relevance to address the problem with new modeling approaches considering the parallel and sequential components of the manufacturing process under equivalent failure in the jobs at each machine. In this paper, we propose a novel Markov chain to model the $N/M/P/F$ permutation flowshop. We analyze the time cost encountered due to $M$ consecutive machine equivalent failures in processing $N$ jobs. We derive new closed form expressions of the completion time (CT) under such setup. We extend the Markov model into its underlying components, providing another new Markov model with processes that don't encounter failures and compare both systems. We provide new insights on job ordering decision rules and new approaches in a set of proposed algorithms that provide novel optimal and heuristic methods that provides optimal or near optimal schedules. We derive closed form expressions that divide per machine CT and per machine processing and waiting times. Further, we provide a novel scheme that proves intimate connections between such time components and the maximum number of rounds per machine that allows optimal utilization of the machines in one CT.

*Note to Practitioners:*
This paper was motivated by the problem of scheduling jobs/processes over machines in a production line or scheduling sequential jobs/processes in a multi-agent system to allow for better utilization of the machines/agents/users/resources. This could apply to problems in manufacturing and automation or other fields of computer science and engineering, and telecommunications. Existing approaches that solve such class of scheduling problems are limited by the size of the problem, and so they depart from providing general or near optimal heuristics for the problem with large size. This paper suggests a new look into the problem using Markov modeling, a probabilistic methodology, that allows for building time models of the process and its underlying components. Utilizing the modeling of the process, we were able to propose new approaches that span between being optimal or near optimal based on the acceptable level of computational complexity. In this paper, we propose novel Absorption Markov chains to derive mathematical models that allow for deriving novel algorithms that searches optimal job ordering/ near-optimal schedules with minimal job/process completion time over machines/agents/users/resources in a permutation flowshop. Then, we propose a new scheme that suggests a new class of joint optimization techniques for minimum completion time-maximum utilization of machines. The models and the solutions provided are applicable to several classes of problems within different applications and fields. Future research will consider the extension of the models and algorithms to machines/agents/users/resources with ad-hoc availability and/or unbalanced failures. Besides, this work provides a milestone towards the selection of optimal time distribution that leads to optimal approaches and optimal decision making for the general optimal scheduling problem.

## I. INTRODUCTION

Optimal scheduling of an $N/M/P/F$ flowshop problem has been addressed extensively in the literature. Since the optimal approach provided by Johnson [1] for the two-machines $N/2/P/F$ flowshop scheduling, several contributions and heuristic approaches to find near optimal job ordering has been proposed. In [2], Palmer provided a heuristic approach based on the notion of slope index. The slope index gives larger weights to jobs that have higher tendency to be processed first. Therefore, the scheduling order is generated based upon a non-increasing order of the slope indexes.

In [3], the authors proposed a heuristic approach for flowshop that relies on Johnson's rule, then creates several schedules from which a best makespan minimizing schedule can be chosen.

In [4], Gupta provides a simple heuristic algorithm which is similar in its theme to Palmer's. In particular, in such approach, a slope index is counted based on accumulative times between consecutive jobs, and the scheduling order is based upon the non-increasing order of the slope indexes. Later, other heuristics based on the previous approaches, using artificial intelligence and optimization techniques have been proposed, the interested reader can refer to an extensive survey on such approaches in [5].

Minimizing the total completion time or the makespan are two main objectives that are used to address optimal or heuristic ordering of jobs in a schedule. Both objectives inherently include the minimization of the waiting times of the machines during their idle times. The simplest case of two-machine flowshop scheduling problem is known to be polynomially solvable when the objective is to minimize makespan, [1]. However, the problem is known to be NP-hard when the performance measure is the total completion time (CT), see [6].

The majority of research on the flowshop problems addresses joint processing and setup times of the machines. In, [7] the authors introduce the concept of separated setup and processing times.Later, [8] highlighted that such separation algorithm might depart from optimality if machine breakdowns are considered. They studied, the problem of scheduling $N$ jobs in two-machine flowshop where the machines are subject to random non-deterministic breakdowns.

In [9] the flowshop scheduling problem through the no-idle, or no-waiting time machines was first to be addressed. The authors main contribution is a polynomial algorithm for solving the two-machines $N/2/P/F$ flowshop scheduling. In [10], the author addressed the general $M$-machine no-idle $N/M/P/F$ flowshop and proposed some heuristics for the makespan objective. Later, in [11], the authors proposed a branch and bound method for the $M$-machine no-idle $N/M/P/F$ with makespan criterion and proved the NP-Hardness of the problem. A set of heuristics has been proposed for the three-machines no-idle $N/3/P/F$ flowshop in [12] and [13].

In [14], the authors proposed new heuristics for the $M$-machine no-idle $N/M/P/F$ that outperform the one proposed in [15] but has higher complexity that is proportional with the number of machines. Most recently, in [16] the authors studied a mixed no-idle flowshop where some machines have the no-idle constraint and others are regular machines. They used a heuristic to construct improved initialization iterated Greedy algorithm. The interested reader can refer to a detailed review on scheduling in flowshops with no-idle machines in [17].

In this paper, we aim to provide a new perspective into the flowshop scheduling problem, we assume the machines are always available, and job setup times are included in the processing. First, we focus on the modeling part of the processing of jobs given that such jobs might fail to be delivered and so it needs to be repeated. Then, we propose novel Markov models characterizing the parallel and sequential parts of processing jobs over different machines. We characterize an underlying structure of hidden states within such Markov model. We, compare such model with failures to a model without failures, and derive new closed form expressions of the completion time, the maximum processing time under parallel processing states, the processing and waiting times per machine, then we propose new approaches for job ordering. We capitalize on the previous analysis to propose new scheme that allows re-usage of machines during their idle times while other machines are running their processes in one completion time. This proposal is of particular relevance not only for machine utilization, but for providing new perspective, new formulations, and new directions to providing solutions to the flowshop scheduling problem with less computational complexity.

The paper is organized as follows, section II introduces the $N/M/P/F$ flowshop scheduling modeling with job failures and the derivation of the completion time, in addition to two new Algorithms that allow for finding optimal job ordering. Section III introduces the $N/M/P/F$ flowshop scheduling modeling without job failures and the derivation of the completion time, per machine processing and waiting times, and propose two new Algorithms that solve the job scheduling in optimal and heuristic ways. We provide also a novel geometrical interpretation of the waiting times with respect to the optimal job ordering via algebraic methods. Section IV introduces a novel scheme that suggests the re-usage of machines within one completion time and derive new closed form expressions of the completion time with and without failures, and characterize upper bounds on the maximum number of rounds machines can be used in one completion time, and characterize the exact number of rounds when no failures exist, then, a new mathematical optimization formulation to the flowshop scheduling problem is proposed. Finally, Section V presents illustrative results and the paper is concluded in Section VI.

## II. $N/M/P/F$ FLOWSHOP SCHEDULING MODELING WITH JOB FAILURES

We consider $N/M/P/F$ an $N$-jobs, $M$-machines, permutation flowshop with machines' failures. The known flowshop with successful processing of all jobs underlies part of a failure success framework. In this section, we address $N/M/P/F$ with machine failures, in the next section we will analyze such flowshop problem considering the success of all jobs. The analysis of both frameworks allow for providing optimal and heuristic approaches to the well-known $NP$-hard problem of minimizing the total completion time.

Absorption Markov chains is an efficient tool that is usually used to model and express the timing of the processes [18]. It has been used in the context of modeling production lines and machines schedules [19], [20]. In [19], such models are used to find the expected value and variance for resources consumed by the process, and derive sufficient and necessary conditions for optimum batch production quantity. [20] used such tool to similarly estimate the material requirements in a production systems under scrapping and reworking. Further, such tool has been very relevant in other fields to address scheduling problems as well [21]. Therefore, and to the best of our knowledge, this work is the first to address the flow-shop scheduling problem using Markov modeling to find optimal and heuristic approaches of the job ordering of schedules on machines.

In particular, in this contribution, we propose an absorption Markov chain that allows for modeling the parallel and sequential parts of the jobs on different machines, given that the sequences of the jobs will be fixed. Figure 3 illustrates the proposed model.

The state in the model corresponds to the state of parallel or single runs of the machines. Figure 3 shows the proposed flowshop machine scheduling model with higher level abstraction of 3 machines example, for illustration purposes, however the model can be extended to the $N$ jobs $M$ machines case. In particular, each machine will run 4 jobs, the state $S_1$ corresponds to the fact that machine $M_1$ is running. The state $S_2$ means that machines $M_1$ and $M_2$ are running in parallel and processing a certain job under a certain time interval that might intersect with the next state if a new machine starts to function, thus, we become at state $S_3$ which means that the machines $M_1$, $M_2$, and $M_3$ are running in parallel and processing a certain job in a specific time interval. The time interval corresponds to the time a number of machines are running in parallel to finish their sequential jobs. The state $S_4$ corresponds to the run of machines $M_1$, $M_2$, and $M_3$ processing a new job. Therefore, two states like $S_3$ and $S_4$

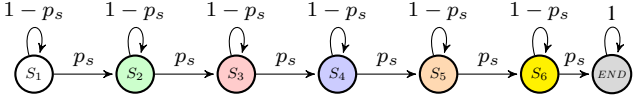

Figure 1. $N = 4$ jobs $M = 3$ machines flowshop scheduling Absorption Markov Model with Number of States $M + N - 1 = 6$, and one absorption state $END$, and probability of successful job delivery is $p_s$ and probability of job failure is $p_f = 1 - p_s$

might correspond to a state where the same machines are running, however, processing new jobs. This identifies the time interval at which machines are processing in parallel, and the time intervals at which are moving sequentially in processing the batches of jobs they need to process. State $S_5$ corresponds to the run of machines $M_2$ and $M_3$, without $M_1$ since it has already been done with its jobs at such point. State $S_6$ corresponds to the run of the last unique job 4 in the last machine $M_3$. Therefore, after such point the Markov chain will be absorbed to the state $END$ with probability 1. However, the movement between states is identified by the probability of successfully finishing a job, called $p_s$, and so moving into a new sequential or parallel state, while if the job at such state failed to be processed with probability $p_f$, no transition will occur until such job is successfully established. Therefore, in a practical system where a production or processing of certain jobs are observed to statistically fail with a percentage of $K/N$ or succeed with $1 - K/N$, each corresponds to the probability of failure or success respectively. For the sake of ease of exploitation of such systems, we assume that the failure and success in the processing of all jobs are equivalent. It is quite interesting to observe that the model need to span over $M + N - 1$ states and absorbed in the $M + N$th state in order to be able to cover all possible states of an $N/M/P/F$ schedule. Worth to note that the one step transition matrix of such absorption Markov chain of Figure 3 can be expressed as:

$$P = \begin{pmatrix} p_{S_1 \to S_1} & p_{S_1 \to S_2} & \cdots & p_{S_1 \to S_{M+N-1}} \\ p_{S_2 \to S_1} & p_{S_2 \to S_2} & \cdots & p_{S_2 \to S_{M+N-1}} \\ \vdots & \vdots & \ddots & \vdots \\ p_{S_{M+N-1} \to S_1} & p_{S_{M+N-1} \to S_2} & \cdots & p_{S_{M+N-1} \to S_{M+N-1}} \\ 0 & 0 & 0 & 1 \end{pmatrix} \quad (1)$$

Where the transition probability $p_{S_k \to S_k} = p_f = 1 - p_s$ defines the failure probability, and the $p_{S_t \to S_k} = p_s, t \neq k$ defines the success probability. The probability of transitioning between one state and another via indirect (non-one-step) paths depends on the machine number and the timing of the preceding jobs in the preceding machines, this will be clarified and discussed in details in section IV. In addition, the last row in the matrix $P$ corresponds to the last state or absorption state in the Markov chain, which we refer to as $END$, at which the whole process is finished with probability 1.

### A. The Completion Time

To derive the completion time for the absorption Markov chain of the $N/M/P/F$ flowshop scheduling, we need to exploit the timing structure that underlies such model. Therefore, for the purpose of analysis we provide a new Markov model that underlies in its timing structure the states exploited in Figure 3. Figure 4 provides a Markov model with each super state in Figure 3 is identified by one or more hidden states. For example, for the case of 3 machines, with each machine having 4 jobs to process, state $\{S(1,1)\} \in S_1$ corresponds to one time interval that is equal to the joint processing and setup time at machine $M_1$, which is equal to $T_{1,1} = T(1,1)$. If the job is processed successfully with probability $p_s$, it will immediately go to a transient state $\{S(1,2), S(2,1)\} \in S_2$ with underlying two processes on machine $M_1$ and machine $M_2$ with joint processing and setup times $T_{1,2} = T(1,2)$ and $T_{2,1} = T(2,1)$ respectively. If the job failed to be processed, the state stays where it is with probability of $p_f$ until its successfully delivered. The cost paid due to failure will be addressed in the timing consideration along the processing paths, until all jobs are processed successfully and the chain is absorbed. It is worth to note that the joint setup and processing time of job $j$ on machine $i$ is represented as $T_{i,j} = T(i,j)$, and the timing of a state $\{S(i,j), S(i+1, j-1), ..., S(M,N)\} \in S_{M+N-1}$ corresponds to the maximum time spent at such state to process all the jobs successfully. So, the state to transit to will not include any repeated sub-states, or in other words, it will not include any repeated job. Further, if under a certain super-state $S_k$ a failure of any process within the set of processes occurs, the additional time cost per job equals $(1 + p_f)T_{i,j}$, and so at such super state, the time spent is equal to the $\max(1 + p_f)T_{i,j} \forall S(i,j) \in S_k$. The completion time based on the absorption Markov chain of Figure 3 for the case of $4/3/P/F$ with $M + N - 1 = 6$ states and one absorption state, can be written as follows:

$$CT^{(1)} = \sum_{k=1}^{7} \frac{T_k}{p_s} \quad (2)$$

With the timing of the state $S_k$ is $T_k = \max_{i,j} T(i,j)_k$ with the time at the $END$ is $T_7 = 0$.

Therefore, we can write the completion time of the $N/M/P/F$ flowshop scheduling with process failures in the following closed form:

$$CT^{(1)} = \sum_{k=1}^{N+M} \frac{T_k}{p_s} \quad (3)$$

where the superscript $.^{(1)}$ means the completion time starting from the start of the job processing at machine $M_1$ at the first state $S_1$ in the Markov chain until all the chain is absorbed finishing all $N$ processes successfully over all $M$ machines, and,

$$T_1 = T(1,1) \quad (4)$$

$$T_2 = \max\{T(1,2), T(2,1)\} \quad (5)$$

$$T_3 = \max\{T(1,3), T(2,2), T(3,1)\} \quad (6)$$

$$T_4 = \max\{T(1,4), T(2,3), T(3,2)\} \quad (7)$$

$$\vdots$$
$$T_\ell = \max\{T(m,\ell), T(m+1,\ell-1), T(m+2,\ell-2), ..., T(M, N-1)\} \tag{8}$$
$$T_{N+M-1} = T(M,N) \tag{9}$$
$$T_{N+M} = 0 \tag{10}$$

Notice that for (8), we can clarify that for a state $S_\ell$ between the first state $S_1$ and the last state $S_{M+N-1}$, the time to absorption from such state includes the maximum of all the times underlying it, so that if $\ell \leq N$, the maximum includes in its set the terms of the time for the machines starting from $m=1$, if $N < \ell \leq N+1$, the maximum includes in its set the terms of all machines excluding $m=1$.

Further, it is clear that the completion time when a failure occurs loads all the times of the chain with the same probability as the division by $p_s$ to all terms $T_k, \forall k = 1,..,M+N$ suggests. Therefore, while we know that searching for an optimal schedule or job ordering that minimizes this completion time is NP-hard, we need to analyze more in depth this completion time when certain setups exist, and so we can reduce the complexity of the problem. Suppose that the completion time $(1-p_f)CT^k = p_f CT^k$, this is a valid assumption when we have a probability of failure $p_f = 0.5$. Of particular relevance to consider such case, since the term $p_f CT^k$ equals zero when $p_f = 0$, however, it adds this amount of time to the normal timing without failures if $p_f > 0$. Thus, under such situation, we derive the total completion time as follows,

$$CT^{(1)} = \sum_{k=2}^{N+M} \frac{p_s^{M+N-k}}{(1-p_s)^{M+N-k+1}} T_{N+M-k+1} \tag{11}$$

This result in (11) is much more interesting to the problem of finding optimal schedule or job ordering. In particular, we can see that the probability of failure introduces different weightings on the timing of the processes, this will be clarified later. And as the minimum of such term corresponds exactly to minimum of (3) due to the fact that the completion time in (3) is double the completion time in (11) since a $50\%$ failure corresponds to doubling the completion time, we can utilize such term to find rules for the decision of optimal job permutations. The following example provides detailed description and better intuition of the effect of failures on the timing of the flowshop machine schedules.

**Example1:** Consider an $4/3/P/F$ flowshop machine scheduling problem with 4 jobs on each machine of the 3 machines. The timing structure follows Markov chains in Figures 3 and Figure 4. Therefore, we utilize the case when $(1-p_f)CT^k = p_f CT^k$ and consider in reversed order the states of the Markov chain, and divide the completion times, starting from each state until the absorption state, as follows:
The completion time starting from state $S_6$ is given by:

$$CT^{(6)} = \frac{1}{p_f} T(3,4) \tag{12}$$

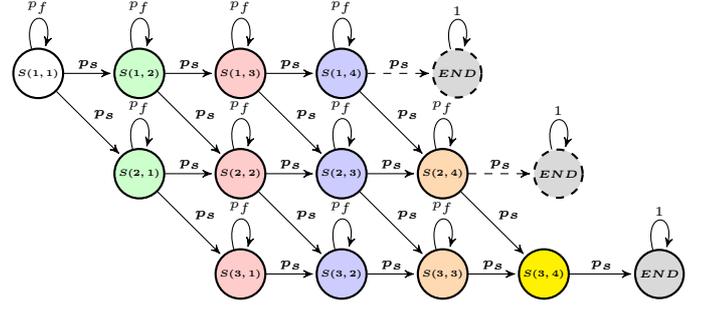

Figure 2. $N=4$ jobs $M=3$ machines flowshop scheduling Absorption Markov Model with Number of States $M+N-1=6$, and one absorption state $END$, and probability of successful job delivery is $p_s$ and probability of job failure is $p_f = 1-p_s$ and underlying processes with timing $T(i,j)$. Each set of colored sub-states corresponds to one superstate in the Absorption Markov chain in Figure 3. The first two dashed circles are absorption states on each machine but not all the process.

The completion time starting from state $S_5$ is given by:

$$CT^{(5)} = \frac{p_s}{p_f^2} T(3,4) + \frac{1}{p_f} \max\{T(2,4), T(3,3)\} \tag{13}$$

The completion time starting from state $S_4$ is given by:

$$CT^{(4)} = \frac{p_s^2}{p_f^3} T(3,4) + \frac{p_s}{p_f^2} \max\{T(2,4), T(3,3)\} + \frac{1}{p_f} \max\{T(1,4), T(2,3), T(3,2)\} \tag{14}$$

The completion time starting from state $S_3$ is given by:

$$CT^{(3)} = \frac{p_s^3}{p_f^4} T(3,4) + \frac{p_s^2}{p_f^3} \max\{T(2,4), T(3,3)\} + \frac{p_s}{p_f^2} \max\{T(1,4), T(2,3), T(3,2)\} + \frac{1}{p_f} \max\{T(1,3), T(2,2), T(3,1)\} \tag{15}$$

The completion time starting from state $S_2$ is given by:

$$CT^{(2)} = \frac{p_s^4}{p_f^5} T(3,4) + \frac{p_s^3}{p_f^4} \max\{T(2,4), T(3,3)\} + \frac{p_s^2}{p_f^3} \max\{T(1,4), T(2,3), T(3,2)\} + \frac{p_s}{p_f^2} \max\{T(1,3), T(2,2), T(3,1)\} + \frac{1}{p_f} \max\{T(1,2), T(2,1)\} \tag{16}$$

The completion time starting from state $S_1$ is given by:

$$CT^{(1)} = \frac{p_s^5}{p_f^6}T(3,4) + \frac{p_s^4}{p_f^3}\max\{T(2,4), T(3,3)\}+$$
$$\frac{p_s^3}{p_f^4}\max\{T(1,4), T(2,3), T(3,2)\}+$$
$$\frac{p_s^2}{p_f^3}\max\{T(1,3), T(2,2), T(3,1)\}$$
$$+ \frac{p_s}{p_f^2}\max\{T(1,2), T(2,1)\} + \frac{1}{p_f}T(1,1) \quad (17)$$

where as stated previously, the superscript $CT^{(1)}$ means the completion time starting from the start of the job processing at machine $M_1$ at the first state $S_1$ in the Markov chain of Figure 3 until all the chain is absorbed at state $END$ with probability 1 finishing all processes successfully. In a similar way to the representation of (11), we can express the total completion time over the $4/3/P/F$ scheduling in (17) as follows:

$$CT^{(1)} = \sum_{k=2}^{7} \frac{p_s^{7-k}}{(1-p_s)^{8-k}} T_{8-k} \quad (18)$$

which simplifies to,

$$CT^{(1)} = \frac{p_s^5}{(1-p_s)^6}T_6 + \frac{1}{(1-p_s)}T_1 + \sum_{k=3}^{6} \frac{p_s^{7-k}}{(1-p_s)^{8-k}}T_{8-k}$$
$$= \frac{p_s^5}{(1-p_s)^6}T(3,4) + \frac{1}{(1-p_s)}T(1,1) + \sum_{k=3}^{6} \frac{p_s^{7-k}}{(1-p_s)^{8-k}}T_{8-k} \quad (19)$$

Where the total completion time is $CT^{(1)}$ starting from the first machine until the chain is absorbed, other completion times $CT^{S(i,j)}$ corresponds to the state at which certain job $j$ started at a certain machine $i$ until the chain is absorbed. Further, its worth to note the weights introduced due to success and failure probabilities in the jobs, we provide the following example to illustrate such concept and to understand its contribution to the decision making of optimal schedules.

**Example2:** Consider a $4/3/P/F$ flowshop scheduling problem. We need to provide the optimal schedule that minimizes the completion time or inherently aims to minimize the makespan and so the waiting or idle times of the machines. Despite the fact that (11) is derived under the assumption that $p_f = 0.5$, however, we can deduce some trends when this value is less than this as well. Let the probability of success and failure in processing the jobs as $p_s = 0.8$ and $p_f = 0.2$, respectively.

In such case, we need to know that the time span will be $M + N - 1 = 6$ corresponding to the steps of the Markov chain in Figure 3. Thus, the series of probabilistic weights generated are $\{\frac{p_s^5}{(1-p_s)^6}, \frac{p_s^4}{(1-p_s)^5}, \frac{p_s^3}{(1-p_s)^4}, \frac{p_s^2}{(1-p_s)^3}, \frac{p_s}{(1-p_s)^2}, \frac{1}{(1-p_s)}\}$ equals $\{5121.375, 1280, 320, 80, 20, 5\}$.

It is worth to observe the accumulated effect of the success and failure probabilities in the processes timing, where as much as we go forward in the processing higher weights are obtained. In particular, assume that the given schedule $j_1, j_2, j_3, j_4$ is the one that allows for minimal total completion time, i.e. the optimal schedule, it is clear in that case that the last step at $S_6$ with the last unique job run on last machine with processing time $T(3,4)$ is provided the highest weight $\frac{p_s^5}{(1-p_s)^6} = 5121$. Similarly, the starting weight at $S_1$ with the first unique job run on first machine with processing time $T(1,1)$ is with least weight $\frac{1}{(1-p_s)} = 5$. Therefore, the main concern stays with the optimal job permutations selection over the intermediate states $S_2, S_3, S_4, S_5$. The following subsection introduces the optimization problem, and proposals to reduce the search space.

### B. Optimal jobs ordering

To define the optimization problem with the objective to minimize the total completion time over all possible job permutations, we can state it mathematically as follows:

$$\min_{\pi(j), \forall j=1,..,N} CT^{(1)} \quad (20)$$

The optimization problem in (20) objective is to find the optimal job permutation that corresponds to the job schedule that reduces the time over all machines. However, a search space with large number of jobs and large number of machines was proven to be an NP-hard problem, as the number of permutation required equals $N!$. Therefore, to reduce the search space its instructive to observe that a straightforward decision of optimal ordering that can reduce the search space over all possible jobs permutations could allow a reduction of $N(N-1)$ operations, or in other words a search can be performed if we perform a search over $(N-2)!$ jobs. This is possible given that we can associate the least weight to least job time in the first machine $M_1$, and associate the most weight to the least timely job at the last machine, therefore, we decide in an optimal way, the first and last job to be processed on the machines. In turn, the optimization problem become as follows,

$$\min_{\pi(j), \forall j=2,..,N-1} CT^{(1)} \quad (21)$$

Algorithm 1 is a new algorithm that explains the evaluation of the completion time and some rules related to optimal ordering of jobs with failures, where as discussed before, we decide for the first and last jobs at all machines, then we search for the optimal order of the intermediate schedules that minimizes the completion time.

However, when the number of jobs or machines are large enough not to make the problem solvable in polynomial time, and so the space at which a large number of operations need to be performed is not feasible to search, we propose a new reduced complexity Algorithm 2 with min-max criterion. In Algorithm 2, we reduce the size of span and we iteratively remove one job or more at each iteration of the intermediate jobs. Thus, the selection of the first and last jobs remains the same as in Algorithm 1, however, the search of the optimal ordering of intermediate jobs is of reduced size over time. Here, we replace the completion time, with a measure that leads to the permutation with the minimum of the maximum time in a similar way to the makespan. In fact, this measure is not exactly the makespan, but a measure of the maximum timing over all machines running parallel processed jobs at a point instant, which corresponds to one state in the absorption Markov chain in Figure 3. Of particular relevance to observe that Algorithm 2 allows two, three, more or less number of machines to contribute in the ordering decision.

---

**Algorithm 1:** Optimal Job Ordering for an $N/M/P/F$ flowshop with job failures

**Inputs:**
Number of Machines: $\mathcal{M}=[1, 2, 3,..., M]$
Number of Jobs: $\mathcal{N}=[1, 2, 3,..., N]$
Time $T_{i,j}$ vector per job $i$, per machine $j$: $[T_{1,i}, T_{2,i}, ..., T_{N,i}]$
probability of success $p_s$ and probability of failure $p_f = 1 - p_s$

**function1:** Generate timing matrix $T$ of size $M \times N$

**function2:** Re-order the timing matrix $T$ of size $M \times M + N - 1$

**function3:** Sort the first machine $M_1$ jobs in increasing order, and select the job with $\min_{1,j_{\forall j=1,..,N}} T_{1,j}$ as first job in the $N/M/P/F$ schedule.

**function4:** Sort the last machine $M_M$ jobs in increasing order, and select the job with $\min_{M,j_{\forall j=1,..,N}} T_{M,j}$ as last job in the $N/M/P/F$ schedule.

**function5:** Generate all possible permutations $\pi(j), \forall j = 2, .., N-1$ find the completion time of each $\pi(j)$ as in (4), (5), ..., to (10)

**function6:** Find the optimal job schedule or optimal permutation that leads to the $\min_{\pi(j)\forall j=2,...,N-1} CT^1$

**Output:** the optimal job schedule $\pi(j)^* = \{j_1, j_2, .., j_{N-1}, j_N\}$.

---

## III. $N/M/P/F$ Flowshop scheduling without failures

To be able to do better analysis and to provide solutions to a general case with or without failures, its worth to analyze the model when no failure exists and compare the cost paid due to failure, large completion time, large waiting times, where the last three are cast basically due to sub-optimal or worst case ordering. In fact, such comparison will allow us to provide further details and much rigorous insights into the solution setup provided in the two previous algorithms.

In principle, the timing matrix that corresponds to the $N$ jobs processed over $M$ machines, will be expressed by a timing matrix $T$ that is of size $M \times M + N - 1$ instead of the usual treatment of the timing of jobs over $M \times N$, allowing for the exploitation of a natural shifting over the time with $M-1$ extra dimensions to be spanned. This is well explained as a concurrent processing of jobs on all machines as

---

**Algorithm 2:** Optimal Job Ordering for an $N/M/P/F$ flowshop with job failures and min-max criterion

**Inputs:**
Number of Machines: $\mathcal{M}=[1, 2, 3,..., M]$
Number of Jobs: $\mathcal{N}=[1, 2, 3,..., N]$
Time $T_{i,j}$ vector per job $i$, per machine $j$: $[T_{1,i}, T_{2,i}, ..., T_{N,i}]$
probability of success $p_s$ and probability of failure $p_f = 1 - p_s$

**function1:** Generate timing matrix $T$ of size $M \times N$

**function2:** Re-order the timing matrix $T$ of size $M \times M + N - 1$

**function3:** Find the optimal job $\pi(j) = j_1, ..., j_N$, with $\pi(j|k)$ is a permutation of $j$ with reduced space $|j - k|$ conditioned on the knowledge of $k$ based on the following steps.

Step1: Schedule or optimal permutation over all machines as follows:

Job $j_1$: Sort the first machine $M_1$ jobs in increasing order, and select the job with $\min_{1,j_{\forall j=1,..,N}} T_{1,j}$ as first job in the $N/M/P/F$ schedule.

Job $j_N$: Sort the last machine $M_M$ jobs in increasing order, and select the job with $\min_{M,j_{\forall j=1,..,N}} T_{M,j}$ as last job in the $N/M/P/F$ schedule.

Step2: Generate all possible permutations $\pi(j), \forall j = 2, .., N-1$. For all jobs except for $j_1$ and $j_N$
Job $j_2$: $\min_{\pi(j|j_1,j_N)\forall j=1,..,N} \max_{j_1,j_2}\{T_{1,j_2}, T_{2,j_1}\}$, the optimal job schedule is done for $first, second$ jobs.

Step3: Generate all possible permutations $\pi(j), \forall j = 3, .., N-2$. For all jobs except for $j_1, j_2$ and $j_N$
Job $j_3$:
$\min_{\pi(j|j_1,j_2,j_N)\forall j=1,..,N} \max_{j_1,j_2,j_3}\{T_{1,j_3}, T_{2,j_2}, T_{3,j_3}\}$,
the optimal job schedule is done for $first, second, and third$ jobs.
repeat steps,
Job $j_{N-1}$:
$\min_{\pi(j|j_1,j_2,..,j_N-2,j_N)\forall j=1,..,N} \max_{j_{N-1},j_N}\{T_{M-1,j_N}, T_{M,j_{N-1}}\}$.

**Output:** the optimal job schedule $\pi(j)^* = \{j_1, j_2, .., j_{N-1}, j_N\}$.

---

columns in the matrix, while sequential processing of jobs in several machines corresponds to the row corresponding to the machine index. Therefore, the timing matrix can be expressed as follows:

$$T = \begin{pmatrix} T_{1,1} & T_{1,2} & \cdots & T_{1,N} & 0 & 0 & 0 & 0 \\ T_{1,1} & T_{2,1} & T_{2,3} & \cdots & T_{2,N} & 0 & 0 & 0 \\ T_{1,1} & T_{2,1} & \vdots & \ddots & \vdots & \vdots & 0 & 0 \\ T_{1,1} & T_{2,1} & T_{3,1} & \vdots & T_{M,1} & T_{M,2} & \cdots & T_{M,N} \end{pmatrix}$$
(22)

Recall that the absorption Markov chains in Figure 3 and

Figure 4 are yet suited to the scenario when no failures exist and so no self loops on top of each state, i.e. $p_f = 0$, and so $p_s = 1$. Its worth to notice that we are not obliged to repeat any job for $(1+p_f)$ of its timing, or in other words, each time will appear once. It is also worth to observe the structure of the timing matrix $T$, where isolating the first and last column, i.e., the column with $T_{1,1}$ and $T_{M,N}$, respectively will reduce the search space when trying to find the optimal job ordering. This conforms with the previous findings where the maximum weight associated to the last job in the last machine $M_M$ assures the choice of the job with minimum time. Further, the repeated appearance of the timing of the first job $T_{1,1}$ suggests (but does not necessitate) minimum time to be associated in the order of the jobs of the first machine $M_1$.

### A. The Completion Time

Similar to the analysis of the flowshop model with failures, for an $N/M/P/F$ with $N$ jobs, $M$ machines permutation flowshop problem, we can write the completion time starting from machine $M_1$ and ending at the last machine $M_M$ as follows,

$$CT^{(1)} = \sum_{j=1}^{M+N-1} \max_{\forall i=1,\ldots,M} \{T(i,j), T(i-1,j+1), T(i-2,j+2), \ldots, T(j,N)\}$$

$$= T(M,N) + T(1,1) +$$
$$\sum_{j=2}^{M+N-3} \max_{\forall i=1,\ldots,M} \{T(i,j), T(i-1,j+1), T(i-2,j+2), \ldots, T(j,N)\} \quad (23)$$

It is clear that the completion time in (23) is exactly the same as the one in (3) when $p_s = 1$, however, we its more feasible to split the terms into $M+N-1$ distinguishable terms. Similar to the framework with failures, the term $T(M,N)$ of the last job $j = N$ in the last machine $M_M$ appears separate, making it possible to associate the last machine job with least processing time as the last in the job schedule, as discussed previously. Additionally, the term $T(1,1)$ of the first job processing $j = 1$ in the last machine $M_1$ appears separate, making it possible to associate the first machine job with a suggested minimal processing time.

Notice that if the jobs are ordered in such a way such that $T(i,j) > T(i-1,j) \forall i = 1, \ldots, M$, at each job $j$, then we can split the general term with sum of maximums in (23) into the following,

$$CT^{(1)} = T(M,N) + \sum_{i=1}^{M-1} T(i,1) +$$
$$\sum_{j=1}^{N-1} \max \{T(M,j), T(M-1,j+1), \ldots, T(j,N)\} \quad (24)$$

Notice that the last term in (24) corresponds to the sum of the times of the first jobs in all $i = 1, \ldots, M-1$, and the second sum corresponds to the maximum times over the last $N-1$ columns in the timing matrix $T$, where the term $T(M,N)$ is unique. Therefore, under this special case, to minimize the total completion time, its important to notice that the term $\sum_{i=1}^{M-1} T(i,1)$ provides a geometrical interpretation to the optimal job schedule. In particular, it enforces the idea that if all other timing paths are parallel to such path over the Markov chain, a minimal completion time can be guaranteed, with a job schedule of optimal order. If this path is longer in time, it will definitely cause larger times to absorption. As a first guess, we might think that we can introduce another new ordering protocol for this special case in conjunction with the ones discussed in Algorithm 1 and Algorithm 2, however, this might be more applicable to the case of non-permutation flowshop $N/M/NP/F$ which is outside the scope of this paper.

Going back to (23), we can write the optimization problem as follows,

$$\min_{\pi(j)\forall j=1,\ldots,N} CT^{(1)} = \min_{\pi(j)\forall j=1,\ldots,N} T(M,N) +$$
$$\min_{\pi(j)\forall j=1,\ldots,N} T(1,1) +$$
$$\min_{\pi(j)\forall j=1,\ldots,N} \sum_{j=1}^{M+N-3} \max_{\forall i=1,\ldots,M} \{T(i,j), T(i-1,j+1), \ldots, T(j,N)\}$$
(25)

Where the optimization problem over all possible permutations in (25) can be solved via the algorithms discussed before. Notice that if the job does not exist i.e. if the sum $j + k > N$, the term $T(i, j+k)$ will be excluded from (29). The following example illustrates the concepts discussed and allows for in depth understanding on the decision making of such heuristic job schedules that might be near optimal.

**Example3:** Consider a $4/3/P/F$ flowshop scheduling problem with no machine or job failures. We need to provide a schedule that minimizes the completion time or inherently aims to minimize the makespan and so the waiting or idle times of the machines.

We will write the total processing time of all jobs in a machine, so unlike the analysis in previous sections, we use a reversed approach. Before, we measure the completion time of each step from the time the machine starts processing until the absorption $END$ state, which means until all other consecutive machines comes after and completes the processing of their jobs. Hence, we provide another approach where we only consider the time each individual machine takes to finish its own jobs accounting for its waiting time due to preceding machines running processes. Later, we will provide the completion time that is in fact the time for the last machine $M_3$ to finish its own processing including all its waiting time until the $M - 1 = 2$ preceding machines $M_1$ and $M_2$ process all their jobs. Its very important to take into account not to double count the times over the counting process of times, as will be shown. We will denote the total processing and waiting time of a machine $i$ by $T_{M_i}$ to avoid confusing it with the total completion time $CT^{(i)}$.

The total processing and waiting time at machine $M_1$ for its 4 jobs is as follows:

$$T_{M_1} = T(1,1) + T(1,2) + T(1,3) + T(1,4) \quad (26)$$

The total processing and waiting time at machine $M_2$ for its

4 jobs is as follows:

$$T_{M_2} = \underbrace{T(1,1)}_{\omega_{M_2,1}} + \underbrace{\max\{T(1,2), T(2,1)\}}_{T(2,1)+\omega_{M_2,2}} + \\ \underbrace{\max\{T(1,3), T(2,2)\}}_{T(2,2)+\omega_{M_2,3}} + \\ \underbrace{\max\{T(1,4), T(2,3)\}}_{T(2,3)+\omega_{M_2,4}} + T(2,4) \quad (27)$$

The total processing and waiting time at machine $M_3$ for its 4 jobs is as follows:

$$T_{M_3} = \underbrace{T(1,1)}_{\omega_{M_3,1}} + \underbrace{\max\{T(1,2), T(2,1)\}}_{T(2,1)+\omega_{M_3,2}} + \\ \underbrace{\max\{T(1,3), T(2,2), T(3,1)\}}_{T(3,1)+\omega_{M_3,3}} + \\ \underbrace{\max\{T(1,4), T(2,3), T(3,2)\}}_{T(3,2)+\omega_{M_3,4}} + \\ \underbrace{\max\{T(2,4), T(3,3)\}}_{T(2,3)+\omega_{M_3,5}} + T(3,4) \quad (28)$$

where $\omega_{M_i,j}$ is the waiting time of machine $M_i$ to start processing due to a certain job processing in previous machine. Notice that for $M_1$, the time span is $N = 4$, for $M_2$, the time span is $N + 1 = 5$, and for $M_3$, the time span is $N + 2 = 6$. This is clear through the number of terms of each $T_{M_i}$, and so the last machine processing and waiting time corresponds to the total completion time of the flowshop with $M + N - 1 = 6$ terms. It follows that the completion time as defined in the previous section from the start of machine $M_1$ until the processing is done over the last machines and the Markov chain is absorbed at the $END$ is $CT^1 = T_{M_3}$. Therefore, we can write it in a general form for the $N/M/P/F$ without failures as $CT^{(1)} = T_{M_M}$. Further, it is clear that (28) matches the expression in (23), with,

$$CT^{(1)} = T(3,4) + T(1,1) + \\ \sum_{j=1}^{4} \max_{\forall i=3,2,1} \{T(i,j), T(i-1, j+1), T(1, j+2)\} \quad (29)$$

Notice that the third term of the sum of maximum durations encapsulates a set of waiting times. Thus, another way to attack the job ordering problem is to try to minimize the waiting times instead of minimizing the total completion time $CT^1$, or the makespan.

### B. Optimal job ordering based on waiting times

In this section, we will address the issue of job ordering not from the perspective of completion time or makespan. We will rely on a setup similar to the example provided in Example3. We can look into the problem from the perspective of waiting times. Worth to note that if $\omega_{M_i,j}$ is minimized the completion time will be directly minimized since the waiting times are basically the main source of large total completion times, and a waiting time $\omega_{M_i,j} = 0$ over the different machines corresponds to parallel processing in the machines, therefore, the optimal job ordering, or optimal schedule remains unnecessary in such case. It follows that, we need to capitalize on the analysis in Example3 to write the waiting times of the general $N/M/P/F$ flowshop machine scheduling problem. The waiting time of the first machine $M_1$ over all jobs $j = 1, ..., N$ equals zero, therefore, we can express it as $\sum_{j=1}^{N} \omega_{M_1,j} = 0$, it follows that the waiting time is more than zero starting from the second machine $M_2$, and given as,

$$\sum_{j=1}^{N} \omega_{M_2,j} = \max\{\sum_{j=1}^{N} \omega_{M_1,j} + \\ \sum_{j=1}^{N} T(1,j) - \sum_{j=1}^{N-1} T(2,j), \sum_{j=1}^{N-1} \omega_{M_2,j}\} \quad (30)$$

Which simplifies to,

$$\sum_{j=1}^{N} \omega_{M_2,j} = \max\{\sum_{j=1}^{N} T(1,j) - \sum_{j=1}^{N-1} T(2,j), \sum_{j=1}^{N-1} \omega_{M_2,j}\} \quad (31)$$

Similarly, the machine $M_{M-1}$ before the last one has a waiting time given as,

$$\sum_{j=1}^{N} \omega_{M_{M-1},j} = \max\{\sum_{j=1}^{N} \omega_{M_{M-2},j} + \\ \sum_{j=1}^{N} T(M-2, j) - \sum_{j=1}^{N-1} T(M-1, j), \sum_{j=1}^{N-1} \omega_{M_{M-1},j}\} \quad (32)$$

The last waiting time in the last machine $M_M$ of the flowshop schedule is given as,

$$\sum_{j=1}^{N} \omega_{M_M,j} = \max\{\sum_{j=1}^{N} \omega_{M_{M-1},j} + \\ \sum_{j=1}^{N} T(M-1, j) - \sum_{j=1}^{N-1} T(M, j), \sum_{j=1}^{N-1} \omega_{M_M,j}\} \quad (33)$$

It is clear that minimizing the waiting time over all machines will minimize the total completion time. It is also clear that minimizing the difference between the waiting times of two consecutive machines to zero or its minimum value, especially that - to a certain limit - waiting is unavoidable, will definitely decrease the total completion time. Therefore, in the following we propose a new Algorithm 3 that measures the waiting times based on the timing of the jobs of each machine. Therefore, in an iterative way, the optimization is to minimize the maximum difference in the waiting times, or in other words the difference of the sums of times of different jobs over two consecutive machines. The job ordering that provides the least minimum over all possible job permutations is the optimal schedule.

It is clear that the processing of Algorithm 3 requires extensive search, despite the fact that its less computationally demanding than the one with minimum completion time. However, to avoid more computational complexity, we will rely on some observations to find a heuristic approach to solve the job ordering problem with the waiting time criterion.

*C. Heuristics for job ordering with waiting time criterion*

In this section, we dig deep inside the equations of the waiting times of each machine. This will allow us to understand further the iterative structure of the problem and its geometrical interpretation. First, we revisit a set of waiting times starting from machine $M_2$ to machine $M_M$ in (31) to (33) respectively. First, we assume that the first term in the $max$ in each equation corresponds to the maximum value. Then, we can re-write the waiting time from machine $M_2$ to $M_M$ in a recursive way, by the substitution of the sum of the waiting times of the preceding machine into the sum of waiting times at the current machine, until the last machines' waiting time. This can be written as,

$$\sum_{j=1}^{N} \omega_{M_2,j} + \sum_{j=1}^{N} \omega_{M_3,j} + ... + \sum_{j=1}^{N} \omega_{M_M,j} =$$
$$\sum_{j=1}^{N} T(1,j) - \sum_{j=1}^{N-1} T(2,j)$$
$$+ \sum_{j=1}^{N} T(2,j) - \sum_{j=1}^{N-1} T(3,j)$$
$$+ .... +$$
$$\sum_{j=1}^{N} T(M-2,j) - \sum_{j=1}^{N-1} T(M-1,j)$$
$$+ \sum_{j=1}^{N} T(M-1,j) - \sum_{j=1}^{N-1} T(M,j) \quad (34)$$

Therefore, as seen in (34) the substitution of the timing of the previous waiting into the current waiting, and taking their sum over all machines, we can notice that there exist two main parts that are much relevant to the minimization problem of the waiting times, as most terms will be canceled keeping only the main timing of the intermediate machines. More particularly, the waiting time for the first and last machine are the most influential machines in this minimization. In light of this, we can propose the following new heuristic approach provided in Algorithm 4 which relies on the minimization of the time differences of the first and last machines in order to decide on the job schedule of the machines.

To make the interpretation of the heuristic approach more clear and to provide insights into geometrical interpretations of the solution, we provide the following example4.

**Example4:** Consider a $4/3/P/F$ flowshop machine scheduling problem. We need to write the timing structure at each machine to find the waiting time at such machine. Given that the waiting time in $M_1$ equals zero, we start with the waiting time at machine $M_2$ given as

**Algorithm 3:** Optimal Ordering for an $N/M/P/F$ flowshop without job failures based on machine waiting times

**Inputs:**
Number of Machines: $\mathcal{M}=[1, 2, 3,..., M]$
Number of Jobs: $\mathcal{N}=[1, 2, 3,..., N]$
Time $T_{i,j}$ vector per job $i$, per machine $j$: $[T_{1,i}, T_{2,i}, ..., T_{N,i}]$

**function1:** Generate timing matrix $T$ of size $M \times N$

**function2:** Re-order the timing matrix $T$ of size $M \times M + N - 1$

**function3:** Sort the first machine $M_1$ jobs in increasing order, and select the job with $\min_{1,j_{\forall j=1,..,N}} T_{1,j}$ as first job in the $N/M/P/F$ schedule.

**function4:** Sort the last machine $M_M$ jobs in increasing order, and select the job with $\min_{M,j_{\forall j=1,..,N}} T_{M,j}$ as last job in the $N/M/P/F$ schedule.

**function5:** Generate all possible permutations $\pi(j), \forall j = 2,..,N-1$

**function6:** Find the job schedule that satisfies the minimum difference in the waiting times of consecutive machines:
$\min_{\pi(j)_{\forall j=1,..,N}} \{ \sum_{j=1}^{N} \omega_{M_M,j} - \sum_{j=1}^{N} \omega_{M_{M-1},j} \} =$
$\min_{\pi(j)_{\forall j=1,..,N}} \max \{ \sum_{j=1}^{N} T(M-1,j) -$
$\sum_{j=1}^{N-1} T(M,j), \sum_{j=1}^{N-1} \omega_{M_M,j} - \sum_{j=1}^{N} \omega_{M_{M-1},j} \}$

**Output:** the job schedule $\pi(j)^* = \{j_1, j_2,..,j_{N-1}, j_N\}$.

$\sum_{j=1}^{N} T(1,j) - \sum_{j=1}^{N-1} T(2,j)$, let us express each element in the first sum as a vector $a = \{T(1,1), T(1,1)+T(1,2), T(1,1)+T(1,2) + T(1,3), T(1,1) + T(1,2) + T(1,3) + T(1,4)\}$ and in the second sum as a vector $b = \{0, T(2,1), T(2,1) + T(2,2), T(2,1) + T(2,2) + T(2,3)\}$. Recall that the zero in $b$ corresponds to the time shift due to waiting of machine $M_2$ for machine $M_1$ to finish the processing of its first job. Therefore, lets define the matrices $A$ and $B$ with the elements of $a$ and $b$ as their diagonal entries, so $A$ reads as:

$$A = \begin{pmatrix} T(1,1) & 0 & 0 & 0 \\ 0 & \sum_{j=1}^{2} T(1,j) & 0 & 0 \\ 0 & 0 & \sum_{j=1}^{3} T(1,j) & 0 \\ 0 & 0 & 0 & \sum_{j=1}^{4} T(1,j) \end{pmatrix}$$
(35)

**Algorithm 4:** Heuristic Ordering approach for an $N/M/P/F$ flowshop without job failures based on machine first-last machines

**Inputs:**
Number of Machines: $\mathcal{M}=[1, 2, 3,..., M]$
Number of Jobs: $\mathcal{N}=[1, 2, 3,..., N]$
Time $T_{i,j}$ vector per job $i$, per machine $j$: $[T_{1,i}, T_{2,i}, ..., T_{N,i}]$

**function1:** Generate timing matrix $T$ of size $M \times N$

**function2:** Re-order the timing matrix $T$ of size $M \times M + N - 1$

**function3:** Sort the first machine $M_1$ jobs in increasing order, and select the job with $\min_{1, j_{\forall j=1,..,N}} T_{1,j}$ as first job in the $N/M/P/F$ schedule.

**function4:** Sort the last machine $M_M$ jobs in increasing order, and select the job with $\min_{M, j_{\forall j=1,..,N}} T_{M,j}$ as last job in the $N/M/P/F$ schedule.

**function5:** Generate all possible permutations $\pi(j), \forall j = 2,..,N-1$

**function6:** Find the job schedule that satisfies the minimum difference in the first machine and last machine times:
$\min_{\pi(j)\forall j=1,..,N} \sum_{j=1}^{N} T(1,j) - \sum_{j=1}^{N-1} T(M,j)$

**Output:** the job schedule $\pi(j)^* = \{j_1, j_2, .., j_{N-1}, j_N\}$.

and $B$ reads as:

$$B = \begin{pmatrix} 0 & 0 & 0 & 0 \\ 0 & T(2,1) & 0 & 0 \\ 0 & 0 & \sum_{j=1}^{2} T(2,j) & 0 \\ 0 & 0 & 0 & \sum_{j=1}^{3} T(2,j) \end{pmatrix} \quad (36)$$

Therefore, the minimal difference between the two matrices is defined by a certain permutation of the eigenvalues of the difference matrix that allows for a minimal Frobenious norm, [22], or in other technical terms that allows smaller determinant, this corresponds geometrically to a smaller volume occupied in space by the difference matrix $A - B$, or by the time vectors associated to each job in each machine.

In a similar way, we can express the waiting times of $M_3$, given as $\sum_{j=1}^{N} T(2,j) - \sum_{j=1}^{N-1} T(3,j)$, let us express each element in the first sum as a vector $\tilde{b} = \{T(2,1), T(2,1) + T(2,2), T(2,1) + T(2,2) + T(2,3), T(2,1) + T(2,2) + T(2,3) + T(2,4)\}$ and in the second sum as a vector $c = \{0, T(3,1), T(3,1) + T(3,2), T(3,1) + T(3,2) + T(3,3)\}$
Therefore, lets define the matrices $\tilde{B}$ and $C$ with the elements of $\tilde{b}$ and $c$ as their diagonal entries, so $\tilde{B}$ reads as:

$$\tilde{B} = \begin{pmatrix} T(2,1) & 0 & 0 & 0 \\ 0 & \sum_{j=1}^{2} T(2,j) & 0 & 0 \\ 0 & 0 & \sum_{j=1}^{3} T(2,j) & 0 \\ 0 & 0 & 0 & \sum_{j=1}^{4} T(2,j) \end{pmatrix} \quad (37)$$

and $C$ reads as:

$$C = \begin{pmatrix} 0 & 0 & 0 & 0 \\ 0 & T(3,1) & 0 & 0 \\ 0 & 0 & \sum_{j=1}^{2} T(3,j) & 0 \\ 0 & 0 & 0 & \sum_{j=1}^{3} T(3,j) \end{pmatrix} \quad (38)$$

With the zero in the first element of matrix $C$ corresponds to the time shift due to waiting of machine $M_3$ for machine $M_2$ to finish the processing of its first job. The minimum of $\tilde{B} - C$ is achieved at a given job schedule permutation that might not necessarily coincide with the one that minimizes $A - B$. But, its clear that the minimum of the total waiting times corresponds to $A - B + \tilde{B} - C$, which makes it clear now that a near optimal heuristic approach could rely on the minimization of $A - C$ for the first and last machines and its corresponding job schedule, as proposed in Algorithm 4.

## IV. COMPLETION TIME WITH AND WITHOUT FAILURES AND RE-USAGE OF MACHINES DURING IDLE TIMES

We propose a novel scheme on top of the proposed model that aims not only to minimize the completion time of one round job processing over multiple machines, however, the proposed scheme allows the machines after finishing their sequential jobs to establish new rounds of job processing. The analysis of providing an optimal schedule can consider such framework of failure, no-failure, and re-usage. So, we consider that, by the time a machine finishes all the jobs associated in the first round of processing a batch of jobs $j = 1,..,N$ it can be reused again. In turn, a combinatorial problem is created, such that by the time a machine is idle, it can be re-utilized for a second round of processing one or more new batches of jobs less than or equal in number to the previous round and within one completion time of all the $M$ machines. For the ease of exploitation, we assume that each new batch has the same number of jobs to process. A first impression could be obtained looking at the timing matrix $T$ in (22) of size $M \times M + N - 1$ where for example machine $M_1$ finishes after processing a batch of $N$ jobs in steps when the process is always successful, which makes it available for re-usage for all the timeslots of zeros in its row, i.e. for $M+N-1-N = M-1$ jobs, less than, equal or more than the length of one batch, depending on the number of machines $M$. Further, machine $M_2$ can process its batch of $N$ jobs in one round and then it can be re-utilized for the right hand side of timeslots with zeros with one timeslot shift until $M_1$ processes its first job, i.e. for $M + N - 1 - 1 - N - 1 = M - 3$ jobs, less than, equal, or more than the length of one batch, depending on the number of machines $M$, and so on and so forth with the same counting process, until the last machine can process one and only one batch of $N$ jobs with almost surely one round. However, to design a job schedule that takes into consideration both aspects, it might be necessary to optimize both the minimum completion time as well as the number of rounds over all machines to allow for the optimum utilization of all machines in one round of completion time over the $M$ machines. In principle, counting the zeros as possible timeslots for the machines to be reused is intuitive, and provides an upper bound on the number of rounds that each machine can be used, however, in reality the timing size is not unified unless all machines has the same jobs' timing and so the problem breaks into no problem. The main issue relies on the fact that each column in $T$ corresponds to different times, where parallel processing of machines is bounded by the maximum time of the machine under such column, making it a problem of a bit more complexity. Further, when there are failures in the machines, considering a given probability (percentage) of failures and success, then a new tailored formulation need to be considered and extended for the completion time. First, we will address the problem for machines encountering failures, and the optimal job selection with an optimal number of rounds of the machines for better utilization. Second, we will address the easier case with the machines not encountering any failures, and propose much more precise way to count the number of rounds it can be run. Therefore, the completion time can be expressed as follows:

$$CT^{S(i,j)} = T_{i,j} +$$
$$\sum_{S'(i,j)=S(i,j+1)}^{S(i,N)} P^{R_i}_{(i,j)\to(i,N+(R_i-1)N)} CT^{S'(i,N+(R_i-1)N)} +$$
$$P^{R_i}_{(i,j)\to(i+1,j)} CT^{S'(i+1,j)} \quad (39)$$

Where $CT^{S(i,j)}$ corresponds to completion time starting at state $S$ at the chain in Figure 3. $R_i$ is the number of rounds or re-processing of jobs when machine $i$ is reused within one completion time of all $M$ machines, therefore, we account for such timing as $T_{i,j}$ since it is not fixed in that case and dependent on the maximum process time at the parallel processing state $S(i,j)$ under state $S$, and we associate the second sum in (39) to count the completion times of the other sequential sub-states $S' = S + 1$ of the same machine after $S$ in Figure 3 until the absorption for this machine with index $i$, in addition to the timing of the completion time of the next processing machine $i+1$ loaded probabilistically with waiting times of the previous machine $i$, and since the finishing of one job $j$ in machine $i$ is associated to the start of the same job $j$ on machine $i + 1$, the time indexes follow the ones in (39). Its quit clear then how the completion time will span $R_i$ steps that appears on the transition matrix between the starting state until a feasible number of rounds $(i,j) \to (i, N+(R_i-1)N)$, this is due to the fact that one round i.e. when $R_i = 1$, it means that we end processing $N$ jobs, if $R_i = 1.5$, it means that we end processing $N + \frac{N}{2}$ jobs in the same machine, so that we don't confuse the concept of a round and how many steps of processing within the transition matrix. The matrix expressing the model in Figure 3 is not anymore suitable to the underlying processes. Therefore, the one step transition matrix of the model proposed in Figure 4 underlying the one in 4, will consider a one step transition matrix per machine, which can be written as follows,

$$P = \begin{pmatrix} 1-p_s & p_s & 0 & \cdots & 0 \\ 0 & 1-p_s & p_s & \cdots & 0 \\ 0 & 0 & \vdots & \ddots & 0 \\ 0 & 0 & \cdots & 1-p_s & p_s \\ 0 & 0 & 0 & 0 & 1 \end{pmatrix} \quad (40)$$

and so the transition matrix over $R_i$ rounds of processing over one machine is written as follows,

$$\left(\prod_{i=1}^{R_i} P\right)_{(i,j)\to(\ell,j+\gamma)} = P^{R_i}_{(i,j)\to(\ell,j+\gamma)}, \quad (41)$$

corresponds to all transition probabilities over the time slots from the initial state $S(i,j)$, until the state at $S'(i, j+\gamma)$ time slot. It follows that we can express (39) in vector form as,

$$[CT^{S(i,j)}] = (I - P^{R_i})^{-1}[T_{i,j}] + P^{R_i}_{(i,j)\to(i+1,j)}[CT^{S(i+1,j)}] \quad (42)$$

Where $[.]$ corresponds to the vector of column elements, $I$ is the identity matrix and $P$ is the transition matrix in (40), and $P^{R_i}_{(i,j)\to(i+1,j)}$ is an element that corresponds to the probability of successfully doing a transition after job $j$ in machine $i$ after $R_i$ rounds back to perform job $j$ in machine $i + 1$. The following example explains in details the proposed scheme and its underlying structure.

**Example5:** Consider a $4/3/P/F$ flowshop machine scheduling problem. We need to write the timing structure at each machine to find the completion time at each machine given that each machine $i \in \mathcal{M} = \{1,2,3\}$ which will be reused for a number of rounds unknown and given by $\mathcal{R} = \{R_1, R_2, R_3\}$. The completion time at machine $M_1$, reads as:

$$\begin{pmatrix} p_s & -p_s & 0 & 0 & 0 \\ 0 & p_s & -p_s & 0 & 0 \\ 0 & 0 & p_s & -p_s & 0 \\ 0 & 0 & 0 & p_s & -p_s \\ 0 & 0 & 0 & 0 & 0 \end{pmatrix}^{R_1} \begin{pmatrix} CT^{S(1,1)} \\ CT^{S(1,2)} \\ CT^{S(1,3)} \\ CT^{S(1,4)} \\ 0 \end{pmatrix}$$
$$= \begin{pmatrix} T(1,1) \\ T(1,2) \\ T(1,3) \\ T(1,4) \\ 0 \end{pmatrix} +$$
$$p_s^{R_1} \begin{pmatrix} p_s & -p_s & 0 & 0 & 0 \\ 0 & p_s & -p_s & 0 & 0 \\ 0 & 0 & p_s & -p_s & 0 \\ 0 & 0 & 0 & p_s & -p_s \\ 0 & 0 & 0 & 0 & 0 \end{pmatrix}^{R_2} \begin{pmatrix} CT^{S(2,1)} \\ CT^{S(2,2)} \\ CT^{S(2,3)} \\ CT^{S(2,4)} \\ 0 \end{pmatrix} \quad (43)$$

The completion time at machine $M_2$ reads as:

$$\begin{pmatrix} p_s & -p_s & 0 & 0 & 0 \\ 0 & p_s & -p_s & 0 & 0 \\ 0 & 0 & p_s & -p_s & 0 \\ 0 & 0 & 0 & p_s & -p_s \\ 0 & 0 & 0 & 0 & 0 \end{pmatrix}^{R_2} \begin{pmatrix} CT^{S(2,1)} \\ CT^{S(2,2)} \\ CT^{S(2,3)} \\ CT^{S(2,4)} \\ 0 \end{pmatrix}$$

$$= \begin{pmatrix} \max\{T(1,2), T(2,1)\} \\ \max\{T(1,3), T(2,2), T(3,1)\} \\ \max\{T(1,4), T(2,3), T(3,2)\} \\ \max\{T(2,4), T(3,3)\} \\ 0 \end{pmatrix} +$$

$$p_s^{R_2} \begin{pmatrix} p_s & -p_s & 0 & 0 & 0 \\ 0 & p_s & -p_s & 0 & 0 \\ 0 & 0 & p_s & -p_s & 0 \\ 0 & 0 & 0 & p_s & -p_s \\ 0 & 0 & 0 & 0 & 0 \end{pmatrix}^{R_3} \begin{pmatrix} CT^{S(3,1)} \\ CT^{S(3,2)} \\ CT^{S(3,3)} \\ CT^{S(3,4)} \\ 0 \end{pmatrix} \quad (44)$$

Finally, the completion time at machine $M_3$ reads as:

$$\begin{pmatrix} p_s & -p_s & 0 & 0 & 0 \\ 0 & p_s & -p_s & 0 & 0 \\ 0 & 0 & p_s & -p_s & 0 \\ 0 & 0 & 0 & p_s & -p_s \\ 0 & 0 & 0 & 0 & 0 \end{pmatrix}^{R_3} \begin{pmatrix} CT^{S(3,1)} \\ CT^{S(3,2)} \\ CT^{S(3,3)} \\ CT^{S(3,4)} \\ 0 \end{pmatrix}$$

$$= \begin{pmatrix} \max\{T(1,3), T(2,2), T(3,1)\} \\ \max\{T(1,4), T(2,3), T(3,2)\} \\ \max\{T(2,4), T(3,3)\} \\ T(3,4) \\ 0 \end{pmatrix} \quad (45)$$

Where the second last term in (45) disappears since the absorption stated time is zero. Additionally, its worth to note that $R_3 = 1$ for the last machine $M_3$.

### A. Optimal scheduling based on optimal number of rounds per machine

An optimal number of rounds $R_i^*$ per machine $i$ corresponds to the maximum number of jobs (or batches if the rounds is greater than or equal $N$ jobs) processed in one completion time of all machines processing. Therefore, the minimum is $R_i = 1_i$, where $1_i$ means that the machine is utilized once in one completion time, this surely applies to the last machine $M$. Therefore, its straightforward to understand that the optimal number of rounds will take the following form with decreasing order $R_1 \geq R_2 \geq ... \geq R_{M-1} \geq 1$, where the last 1 corresponds to $1_M$.

An optimal number of rounds can be though of as a design and optimization criterion to allow for better machine utilization, and thus less idle times. However, we can define our optimal scheduling over multiple rounds of job processing over all machines,

$$\min_{R_1,...,R_M} CT^{S(i,j)} = \min_{R_1,...,R_M} T_{i,j} +$$

$$\sum_{S'(i,j)=S(i,j+1)}^{S(i,N)} \min_{R_1,...,R_M} P^{R_i}_{(i,j) \to (i,N+(R_i-1)N)} CT^{S'(i,N+(R_i-1)N)} +$$

$$\min_{R_1,...,R_M} P^{R_i}_{(i,j) \to (i+1,j)} CT^{S'(i+1,j)} \quad (46)$$

Given that the minimization need to be constrained to the completion time of all machines in one round i.e. subject to $CT^1$ is measured and minimal, therefore, we can combine the objective of minimizing the completion time to find an optimal permutation of the job schedules and a maximum number of machine utilization. Notice that solving such a problem is very complex, and so we can think about some feasible methods to count such optimal rounds. In fact, the maximum number of rounds subject to a given completion time should correspond to the optimal job ordering. So, one way or another, the problem can be solved optimally or by heuristic approaches.

### B. Number of Rounds for Machines without Failures

The problem might become more feasible when no failure encountered by any machine. In such case, we can measure the time the machine can be utilized as the difference between the completion time (starting from the time the machine started processing) and the job processing time at this machine. For, example, if we have $\mathcal{M} = 1, ...M$ machines with $\mathcal{N} = 1, .., N$ batch of jobs each. The completion time is given by $CT^1$ as given in (23) and the total job processing and waiting time (zero for $M_1$) of $N$ jobs at machine $M_1$ is given by $T_{M_1}$ as stated in (26). It follows that the duration that we can reuse machine $M_1$ within this completion time is equal to $\tau_1 = CT^1 - T_{M_1}$. Therefore, machine $M_2$ will be waiting for machine $M_1$ not only for the first waiting time counted within the total job processing time $T_{M_2}$, but, it will require another waiting time for the first job of the second round in machine $M_1$. Therefore, in one completion time, the duration that we can reuse machine $M_2$ equals $\tau_2 = CT^1 - T_{M_2} - R_1 T(1,1)$, and the duration that machine $M_3$ can be reused equals $\tau_3 = CT^1 - T_{M_3} - R_1 T(1,1) - R_2 T(2,1)$, and so on and so forth until the last machine $M_M$ which can be utilized only once time where $CT^1 = T_{M_M}$. Utilizing this framework, it follows that we can write the number of rounds $R_i$ that the machines $i = 1, ..., M$ can be used as follows, For machine $M_i$, the number of rounds $R_i$ or machine utilization or machine reuse, is given by:

$$R_i = 1 + \frac{\tau_i}{T_{M_i}} \quad (47)$$

where the 1 in the right hand side corresponds to the first round of processing one batch of jobs, and the fraction $\frac{\tau_i}{T_{M_i}}$ corresponds to the re-usage time that the machine will be re-utilized, if $\frac{\tau_i}{T_{M_i}} < 1$ then the machine will process partially few more jobs within one batch of $N$ jobs, if $\frac{\tau_i}{T_{M_i}} > 1$ it means that it will be re-utilized for processing one round of batches or more. The following example explains the iterative method of finding the number of rounds that each machine can be reused with.

**Example6:** To illustrate the way the number of rounds are counted for a $N/4/P/F$ flowshop, with $N$ jobs and 4 machines, we find them in an iterative way starting from, the number of rounds $R_1$ for the first machine $M_1$, as follows:

$$R_1 = \frac{CT^1}{T_{M_1}} = 1 + \frac{\tau_1}{T_{M_1}} \quad (48)$$

The number of rounds $R_2$ of the second machine $M_2$ are,

$$R_2 = \frac{CT^1}{T_{M_2}} - \frac{R_1 T(1,1)}{T_{M_2}} = 1 + \frac{\tau_2}{T_{M_2}} \quad (49)$$

The number of rounds $R_3$ of the third machine $M_3$ are,

$$R_3 = \frac{CT^1}{T_{M_3}} - \frac{R_1 T(1,1)}{T_{M_3}} - \frac{R_2 T(2,1)}{T_{M_3}} = 1 + \frac{\tau_3}{T_{M_3}} \quad (50)$$

Finally, the number of rounds $R_4$ of the forth machine $M_4$ is,

$$R_4 = \frac{CT^1}{T_{M_4}} - \frac{R_1 T(1,1)}{T_{M_4}} - \frac{R_2 T(2,1)}{T_{M_4}} - \frac{R_3 T(3,1)}{T_{M_4}} = 1 \quad (51)$$

Where $R_4 = 1$ almost surely since the last machine $M_4$ is utilized only once in one completion time as discussed before.

## V. SIMULATION RESULTS

We shall now present a set of illustrative results that cast further insights into the problem and the solutions provided. First, we will analyze the problem generally with respect to the number of machines, number of jobs and machine failure and success. Our benchmark is the optimal schedule over all possible permutation. Second, we evaluate our proposed algorithms in comparison to the benchmark. Third, we will evaluate the proposed scheme with machine re-usage, where the number of rounds that can be used within one completion time will be measured. Therefore, the main key performance measure is the completion time, however, we will rely on the machines processing plus setup times to measure the number of rounds iteratively.

### A. Benchmark Approach

Table I
JOB TIMING PARAMETERS OF A $10/7/P/F$ PERMUTATION FLOWSHOP

| $Jobs$ | $j_1$ | $j_2$ | $j_3$ | $j_4$ | $j_5$ | $j_6$ | $j_7$ | $j_8$ | $j_9$ | $j_{10}$ |
|---|---|---|---|---|---|---|---|---|---|---|
| $M_1$ | 3 | 6 | 2 | 1 | 2 | 3 | 4 | 5 | 3 | 0.3 |
| $M_2$ | 0.8 | 4.5 | 1 | 0.5 | 2 | 3 | 4 | 2 | 6 | 0.1 |
| $M_3$ | 1 | 2 | 3 | 4 | 7 | 2 | 4 | 5 | 8 | 1.4 |
| $M_4$ | 2 | 0.9 | 1 | 0.5 | 7 | 4 | 3 | 5 | 7 | 1.4 |
| $M_5$ | 0.1 | 0.9 | 5 | 0.5 | 7 | 4 | 2 | 5 | 8 | 6 |
| $M_6$ | 2 | 4 | 1.1 | 1.5 | 7 | 8 | 3 | 5 | 7 | 1.4 |
| $M_7$ | 0.1 | 0.9 | 5 | 0.5 | 6 | 4 | 9 | 5 | 2 | 6 |

First, we consider as a benchmark optimal approach, to evaluate the proposed algorithms. In the benchmark approach, we measure the completion time over all possible permutations and define the optimal schedule or optimal job ordering by the one that is associated to the minimum completion time. However, we define the worst schedule or worst job ordering as the one that is associated to the maximum completion time. Figure 3, Figure 4, and Figure 5 illustrate the time distribution of the jobs over 7, 3, and 2 machines with 10 jobs for each, and the timing is provided in Table I. It is clear how the job ordering affects the completion time $CT^1$ in a significant way making the difference in the completion time between the optimal schedule with $\min CT^1$ and the worst schedule with $\max CT^1$ quite very far if the number of machines or jobs grow. Additionally, when the number of machines are 2 (Figure 5), it is not clear how the time should be allocated in the ordering process, therefore, going into 3 machines (Figure 4), 7 machines (Figure 3) or more, the distribution of the optimal order become more aligned to the selection of the job with minimal timing on the first machine, and the job with minimal timing on the last machine, as our proposed algorithms suggest. Additionally, looking further in depth, we can see that the worst ordering is associated with a distribution that gives maximum timing to first job on the first machine and last job on the last machine, which makes our decision rules defined in the proposed algorithms much more significant. Further, the intermediate jobs follow a similar trend in the optimal ordering, however, its very hard to define a specific rule over them and so we keep the permutations to exist over all of such jobs. Therefore, it is easy to conclude that in such framework, there exist a time distribution that can decrease the total completion time of the process. Of particular interest for future research is to find an optimal time distribution that is a global minimizer of the completion time. Such distribution defines what an optimal schedule or job ordering will be.

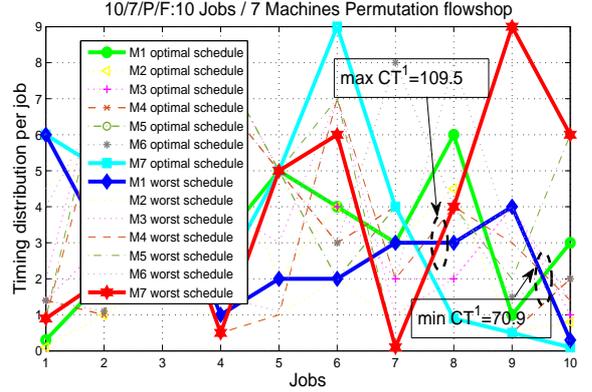

Figure 3. Timing distribution per job versus the jobs in $10/7/P/F$ flowshop under optimal and worst completion times.

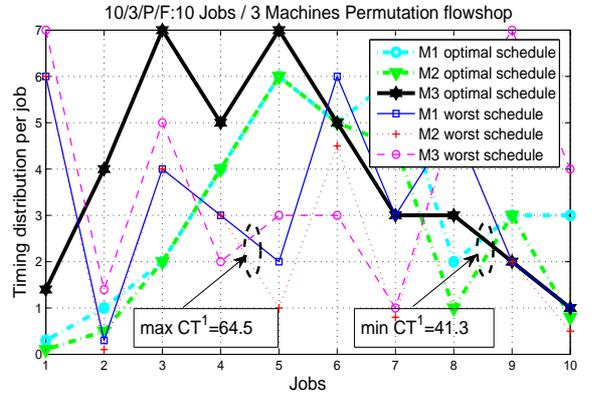

Figure 4. Timing distribution per job versus the jobs in $10/3/P/F$ flowshop under optimal and worst completion times.

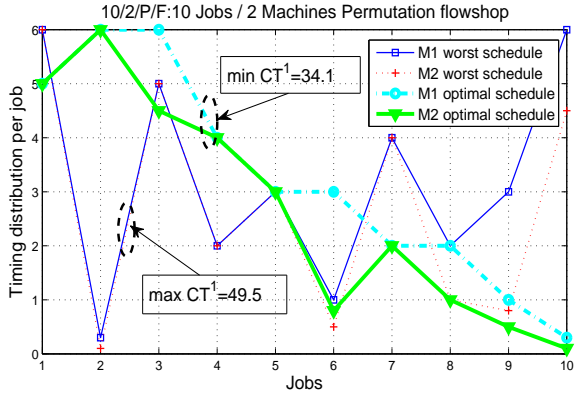

Figure 5. Timing distribution per job versus the jobs in $10/2/P/F$ flowshop under optimal and worst completion times.

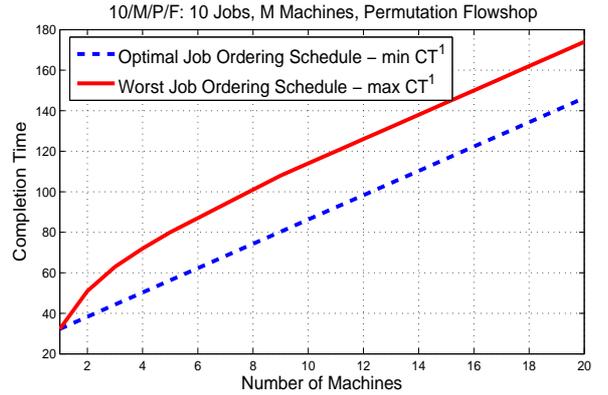

Figure 6. Completion time versus number of machines: comparison for optimal and worst schedules with equivalent job times

## B. Effect of the number of machines

To obtain some intuitions about the effect of the number of machines in the job ordering and completion time. We consider that all machines have $N$ jobs to process with equivalent job processing times across the jobs and not per job, and compare the optimal schedules to the worst schedules in such case. Figure 6 illustrates the minimum completion time with optimal job ordering and the maximum completion time with worst job ordering versus the number of machines. Clear is the increase in the completion time with respect to the increase in the number of machines. More interesting is to notice that the increase in the minimal completion time follows a linear behavior. Such behavior suggests that when the machines have similar number of jobs with equivalent times across the jobs, it is sufficient to design the optimal schedule using few number of machines rather than using all the machines. Such optimal schedule will be equivalent even if more machines are added to the system. Such observation decreases the complexity in the searching process in a significant way. Further, its worth to notice that roughly after 4 machines, the completion time under worst schedules follows a linear behavior as well, and increases with a fixed distance from the one encountered under optimal schedules. This behavior is trivial given that the comparison is held under fixed number of jobs. Therefore, in the next subsection we will consider one more layer of complexity to the problem, which is the effect of the number of jobs.

## C. Effect of the number of jobs

According to the previous analysis, there is a linear increase in the completion time when the machines have the same number of jobs with same timing across the jobs. Therefore, we consider that all machines have $N$ jobs to process with equivalent job processing times across the jobs and not per job. We try to evaluate the behavior of the completion time starting from one job per machine with equivalent times, and inserting further jobs that has different times to the previous one, but equivalent across the machines. Figure 7 illustrates the completion time versus the number of jobs over all machines under optimal and worst schedules for the case of 3, 5, and 10 machines. It is clear that there yet exist intervals over which the completion time increases almost linearly with the increase in the number of jobs. Further, for a small number of jobs the difference between the completion times of optimal and worst schedules is small, however, as the number of inserted jobs increases, the gap becomes wider, making it a more complex problem to solve as much as the number of the parameters increases. Therefore, its worth to add further layers of complexity to the problem and to study the effect of job failures in the completion times.

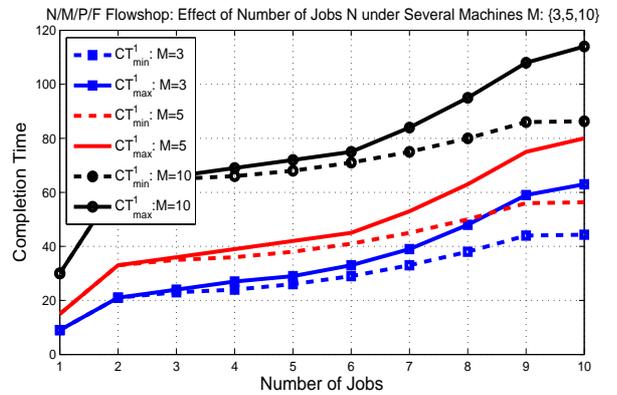

Figure 7. Completion time versus number of Jobs: comparison for optimal and worst schedules with equivalent job times inserted

## D. Effect of the probability of success and failure

As mentioned in the modeling part, a success probability $p_s = 1$ that is associated with a failure probability of $p_f = 1 - p_s = 0$ usually weights the timing process over all machines equivalently. However, a known failure that is roughly constant over all machines [1] will associate different weights to the timing of the process. This can be easily checked by simulation, and via the analytical framework provided. However, when $0 < p_s < 1$ and $p_s \neq 0.5$, there exist job ordering strategies that are different than the ones here. This is a future research work of potential interest. However, when $p_s = 0.5$, the optimal schedules matches those when $p_s = 1$. Figure 8 illustrates the completion time with respect to the probability of success, it is clear that the time increases almost linearly at high probability of success region between 0.9-1, however, when the success rate is less the completion time increases in an exponential way. Figure 9 illustrates a reciprocal behavior to the previous one with respect to the probability of failure, where less machine failures takes the completion times into linear regions. The general aspects of the problem has been analyzed previously. The following will consider the evaluation of the proposed algorithms and the machine re-usage scheme to understand the potential of each solution with respect to the optimal one.

---
[1]The assumption of constant probability of failure over all machines is held for ease of exploitation and to provide insights about the main optimal scheduling problem of machines encountering no failures.

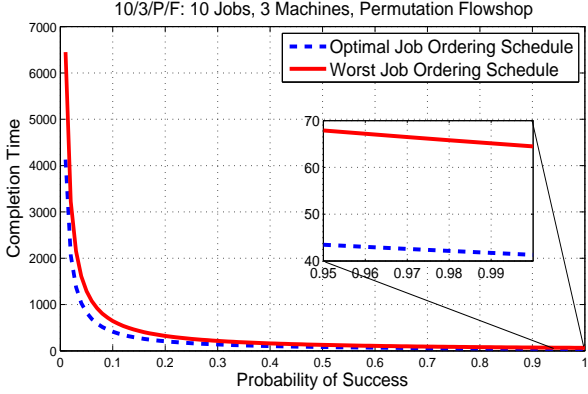

Figure 8. Completion time versus the probability of success with 10/3/P/F

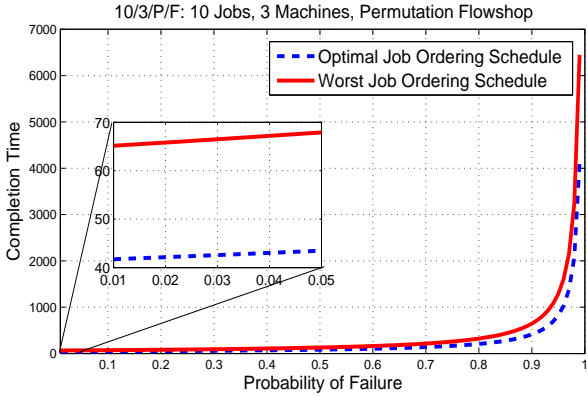

Figure 9. Completion time versus the probability o failure with 10/3/P/F

### E. Performance analysis of the proposed algorithms

Since all the proposed algorithms have shared steps with Algorithm 1, we will first analyze Algorithm 1 in details. Then, we will provide a summary of the performance results of the other algorithms Algorithm 2, Algorithm 3, and Algorithm 4 accordingly in comparison to the benchmark optimal approach.

We evaluate our proposed Algorithm 1 with respect to the optimal benchmark in terms of minimum and maximum completion times. The results for a $10/3/P/F$ flowshop with the timing of the jobs in the three machines are provided in the first three rows of the Table I. The following are the obtained parameters using Algorithm 1:

The minimum completion time=40.3

The maximum completion time=55.5

The optimal and worst schedules correspond respectively to job optimal ordering and job worst ordering, given that the permutation in Table I corresponds to $[j_1\ j_2\ j_3\ j_4\ j_5\ j_6\ j_7\ j_8\ j_9\ j_{10}]$. Therefore,

The optimal schedule using Algorithm 1 is : $[j_{10}\ j_4\ j_5\ j_9\ j_7\ j_2\ j_8\ j_3\ j_6\ j_1]$

The worst schedule using Algorithm 1 is : $[j_{10}\ j_8\ j_7\ j_6\ j_3\ j_2\ j_9\ j_4\ j_5\ j_1]$

Thus, the optimal job timing and the worst job timing are summarized, respectively, in Table II and Table III.

Table II
OPTIMAL JOB TIMING PARAMETERS OF A $10/3/P/F$ PERMUTATION FLOWSHOP USING ALGORITHM 1

| $Jobs$ | $j_{10}$ | $j_4$ | $j_5$ | $j_9$ | $j_7$ | $j_2$ | $j_8$ | $j_3$ | $j_6$ | $j_1$ |
|---|---|---|---|---|---|---|---|---|---|---|
| $M_1$ | 0.3 | 1 | 2 | 3 | 4 | 6 | 5 | 2 | 3 | 3 |
| $M_2$ | 0.1 | 0.5 | 2 | 6 | 4 | 4.5 | 2 | 1 | 3 | 0.8 |
| $M_3$ | 1.4 | 4 | 7 | 8 | 4 | 2 | 5 | 3 | 2 | 1 |

Table III
WORST JOB TIMING PARAMETERS OF A $10/3/P/F$ PERMUTATION FLOWSHOP USING ALGORITHM 1

| $Jobs$ | $j_{10}$ | $j_4$ | $j_5$ | $j_9$ | $j_7$ | $j_2$ | $j_8$ | $j_3$ | $j_6$ | $j_1$ |
|---|---|---|---|---|---|---|---|---|---|---|
| $M_1$ | 3 | 5 | 4 | 3 | 2 | 6 | 3 | 1 | 2 | 5 |
| $M_2$ | 6 | 2 | 4 | 3 | 1 | 4 | 6 | 0.5 | 2 | 2 |
| $M_3$ | 8 | 5 | 4 | 2 | 3 | 2 | 8 | 4 | 7 | 5 |

Moreover, the benchmark providing optimal schedules over all possible job permutations provides the following results:

The minimum completion time=40.3

The maximum completion time=60.5

The optimal and worst schedules correspond respectively to job optimal ordering and job worst ordering, given that the permutation in Table I corresponds to $[j_1\ j_2\ j_3\ j_4\ j_5\ j_6\ j_7\ j_8\ j_9\ j_{10}]$. Therefore,

The optimal schedule using Algorithm 1 is : $[j_{10}\ j_4\ j_5\ j_9\ j_7\ j_2\ j_8\ j_3\ j_6\ j_1]$

The worst schedule using Algorithm 1 is : $[j_9\ j_{10}\ j_6\ j_7\ j_5\ j_3\ j_1\ j_2\ j_4\ j_8]$

In a similar way, we can see that the optimal schedule coincide with the one obtained using Algorithm 1. Therefore, the optimal job timing follows the one in Table II. However, the worst job timing is shown in Table IV.

Comparing Algorithm 1 with the benchmark, its worth to notice that:

1. In the optimal schedule case with minimal completion time, we see that for the case of $10/3/P/F$, Algorithm 1

Table IV
WORST JOB TIMING PARAMETERS OF A $10/3/P/F$ PERMUTATION
FLOWSHOP WITH BENCHMARK: FULL SEARCH OVER ALL PERMUTATIONS

| Jobs | $j_9$ | $j_{10}$ | $j_6$ | $j_7$ | $j_5$ | $j_3$ | $j_1$ | $j_2$ | $j_4$ | $j_8$ |
|---|---|---|---|---|---|---|---|---|---|---|
| $M_1$ | 3 | 0.3 | 3 | 4 | 2 | 2 | 3 | 6 | 1 | 5 |
| $M_2$ | 6 | 0.1 | 3 | 4 | 2 | 1 | 0.8 | 4.5 | 0.5 | 2 |
| $M_3$ | 8 | 1.4 | 2 | 4 | 7 | 3 | 1 | 2 | 4 | 5 |

completion time is exactly equal to the one obtained via the benchmark $\min CT^1 = 40.3$.

2. The optimal schedule obtained by Algorithm 1 matches the one by the benchmark with full span over all $N! = 10! = 3628800$ permutations, only by using $N! - (N-2)!$ number of permutations, allowing for a saving of $(N-2)! = 3588480$ operations in this example.

3. In worst case condition, the maximum time obtained via Algorithm 1 $\max CT^1 = 55.5$ is less than the one with the benchmark with full span over all possible permutations $\max CT^1 = 60.5$. This proves that the selection of the first and last jobs to be the minimum of the first and last machines respectively, is optimal and completion time minimizer under any random distribution (ordering) of the timing of intermediate jobs.

Table V provides a complete summary and comparison of the proposed algorithms with respect to the optimal benchmark in terms of minimum and maximum completion times, for different permutation flowshops, and under equivalent job timing (EJT).

Table V
COMPARISON BETWEEN THE PROPOSED ALGORITHMS AND THE BENCHMARK

| N/M/P/F | Parameters | Benchmark | Alg 1 | Alg 2 | Alg 3 | Alg 4 |
|---|---|---|---|---|---|---|
| 3/10/P/F | $\min CT^1$ | 40.3 | 40.3 | 47.3 | 43.8 | 44.8 |
| | $\max CT^1$ | 60.5 | 55.5 | 47.3 | 52.3 | 52.3 |
| | Number of permutations | 10! | 8! | 1 | 8! | 8! |
| 4/10/P/F | $\min CT^1$ | 46.3 | 46.8 | 55.8 | 54.3 | 54.8 |
| | $\max CT^1$ | 72.5 | 65.5 | 55.8 | 62.3 | 62.3 |
| | Number of permutations | 10! | 8! | 1 | 8! | 8! |
| 4/10/P/F with EJT | $\min CT^1$ | 47.3 | 47.3 | 47.3 | 49.3 | 49.3 |
| | $\max CT^1$ | 63 | 63 | 47.3 | 52.3 | 54.3 |
| | Number of permutations | 10! | 8! | 1 | 8! | 8! |

It is clear that we can obtain a near optimal performance with Algorithm 1 that almost matches the optimal benchmark with two extra gains: first, is the less computations that need to be done with $(N-2)!$ instead of $N!$. The second gain is that the maximum completion time is always less than that of the benchmark since we always enforce the tails of the time distribution to be our optimal selection and so less completion times are encountered. Further, that from an implementation perspective Algorithm 2 can be performed over one permutation without the need to recursively regenerate the permutation, thus, we can see worst performance in terms of minimal completion time within all. Algorithm 3 which depends on the minimization of waiting times provides relevant results that is sometimes near to the optimal, however, we yet have to perform similar computations of $(N-2)!$ with less performance. Additionally, Algorithm 4 provides near optimal performance that is similar to Algorithm 3, only by using the timing of first and last machines saving also $(N-2)$ computations in the algorithm and with same number of permutations $(N-2)!$ over the first and last machines jobs. This casts further insights about the importance of the first and last machines timing in the design of optimal distributions that control such tails in the distributions while minimize the variance of the process, given that the intermediate jobs are spread around the mean of the process, and such variance decreases by decreasing the distance between the waiting times, as the geometrical interpretation of this algorithm explains. To establish the previous observations, we can see Table VI which provides a compact summary of the algorithms that rely on the minimal difference in the waiting times between machines; Algorithm 3, and Algorithm 4 for two cases, excluding first and last jobs (EFLJ) as proposed, and including all jobs and their permutations (IAJ). It is quite clear that for an $N/M/P/F$ flowshop scheduling problem, including all permutations might not necessarily provide better solutions than the proposed exclusion of first and last jobs after deterministic selection. This is due to the fact that a minimal difference between waiting times might not necessarily correspond to minimal waiting time, therefore, if the distribution of the jobs is optimal or near optimal it provides better completion times.

Table VI
COMPARISON BETWEEN ALGORITHMS WITH WAITING TIME CRITERION

| N/M/P/F | Parameters | Alg 3 (EFLJ) | Alg 3 (IAJ) | Alg 4 (EFLJ) | Alg 3 (IAJ) |
|---|---|---|---|---|---|
| 3/10/P/F | $\min CT^1$ | 43.8 | 44.8 | 44.8 | 48.8 |
| | $\max CT^1$ | 52.3 | 52.4 | 52.3 | 52.2 |
| | Number of permutations | 8! | 8! | 8! | 8! |
| | Minimum waiting time | 3 | 1.3 | 1 | 0.3 |
| 4/10/P/F | $\min CT^1$ | 54.3 | 50.3 | 54.8 | 55.8 |
| | $\max CT^1$ | 62.3 | 65.5 | 62.3 | 65.6 |
| | Number of permutations | 8! | 8! | 8! | 8! |
| | Minimum waiting time | 3 | 3.3 | 2 | 0.3 |
| 5/10/P/F | $\min CT^1$ | 53.3 | 56.3 | 55.3 | 62 |
| | $\max CT^1$ | 55.3 | 62 | 61.3 | 65.3 |
| | Number of permutations | 8! | 8! | 8! | 8! |
| | Minimum waiting time | 7.3 | 7.3 | 2 | 0.3 |

Finally, we can see through such comparison between an optimal approach used as benchmark and the proposed algorithms that there is a clear tradeoff between the complexity and the optimal solution. Moreover, the completion time as a performance measure provides better solutions to the job scheduling problems than minimizing the difference in waiting times, for the same or less computational complexity.

*F. Performance analysis of the proposed machine re-usage scheme*

In this subsection, we provide analysis to the proposed scheme, where we will make it clear how we can utilize the machines allowing for no-idle times, and so we allow for a re-usage structure of the machines in one completion time. First, we illustrate as a basic component to the measurement process of the rounds of re-usage of machines, that is the job processing time over each machine. Figure 10 shows the job processing time of 3 machines, each with a batch of 10 jobs to process, i.e. $10/3/P/F$ flowshop, versus the number of permutations which equals $10! = 3628800$. We can clearly see how the processing time gradually increases with the sequence of the machines due to the waiting of one machine for the other previous job processing, we can also see how at a certain permutation $\pi(j)$, there exist a minimum completion time that corresponds to the value of the job processing time of the last machine $\min CT^1 = \min T_{M_3}(\pi^*(j))$, at which a job schedule or job ordering is optimal. Next, we identify the number of rounds that we can iteratively count on each machine. Such rounds will correspond to the number of times that a machine can be utilized in one completion time. For the same example a simple calculation following the explanation in the previous sections lead to a set $\mathcal{R}$ : $\{R_1 = 1.3754, R_2 = 1.2426, R_3 = 1\}$. Figure 11 illustrates the number of rounds each machine can be utilized under different number of machines. Its quite clear the exponential decay in the number of rounds that saturates at level 1, the minimal number of rounds a machine can perform (one batch of jobs) in one minimal completion time. It means that if the completion time is not optimal some machines might be re-used more, however, the main target of introducing the number of rounds is to see how we can maintain minimal completion time via optimal schedules in the optimization problem jointly with the number of rounds to find the optimal job schedules.

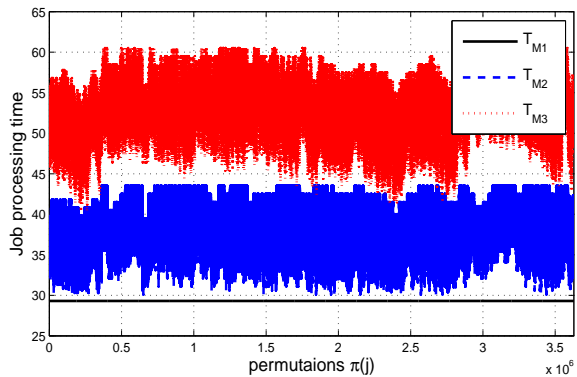

Figure 10. Job processing times versus the permutations for 10/3/P/F

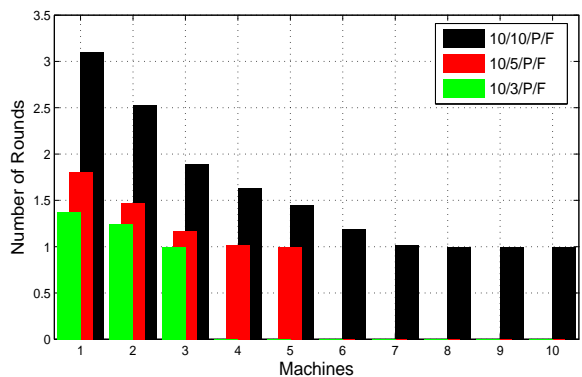

Figure 11. Number of rounds versus the machines

## VI. Conclusion

This paper addresses the permutation flowshop machines scheduling problem. We propose novel models that allow for an in depth study of the timing of the process, and allows for proposals of optimal and heuristic algorithms that provide optimal or near optimal schedules in terms of minimal completion time. The proposed framework is enriched with a novel scheme that allows for machine re-usage, which besides the minimization of the completion time, allows for using machines during idle times, so that no idle times are allowed. Further, this scheme opens dimensions for novel optimization methods with less complexity to attack the problem. Additionally, it allows not only for finding optimal schedules under machine re-usage with success or fixed failures, but it also provides optimal schedules on job-demand by assigning least favorable and most favorable jobs to machines when a certain re-usage pattern is needed. We provide fresh insights about the optimal distribution of the job timing over machines in the tails of the process: the first machine and the last one. Future research directions will consider: First, the analysis of optimal distributions for the whole timing process in the flowshop general scheduling problem. Second, extension of the proposed frameworks to the problem of machines encountering different failure probabilities, thus, new solutions need to be proposed capitalizing on the ones proposed here. Third, finding optimal schedules that minimize the completion time or waiting times subject to a given number of machine rounds of use and re-usage is foreseen.

## VII. Acknowledgment

This work was partially supported by project "NORTE-07-0124-FEDER-000057", under the North Portugal Regional Operational Programme (ON.2 - O Novo Norte) and the National Strategic Reference Framework through the European Regional Development Fund, and by national funds through Fundação para a Ciência e a Tecnologia (FCT).